\documentclass[11pt]{article}
\usepackage{srcltx}
\usepackage{amsmath}
\usepackage{amssymb}
\usepackage{amsfonts}
\usepackage{latexsym}
\usepackage{textcomp}
\usepackage{appendix}
\usepackage{multirow}
\usepackage{booktabs}
\usepackage{array}
\usepackage{marvosym}
\usepackage [ansinew] {inputenc}
\usepackage{graphics}
\usepackage{graphicx}
\title{Data manipulation detection via permutation information theory quantifiers \footnote{Preprint presented at XVIII Conference on Nonequilibrium Statistical Mechanics and Nonlinear Physics MEDYFINOL 2014. Maceió (AL), Brazil.}}

\author{Aurelio Fern\'andez Bariviera\\ 
\footnotesize{Department of Business, Universitat Rovira i Virgili, Av. Universitat 1, 43204 Reus, Spain} \\ \footnotesize{\ttfamily aurelio.fernandez@urv.net} \and M. Bel\'en Guercio \\ \footnotesize{Instituto de Investigaciones Económicas y Sociales del Sur, Universidad Nacional del Sur, Argentina} \\
\footnotesize{Consejo Nacional de Investigaciones Cient\'ificas y T\'ecnicas (CONICET), C1033AA5, Buenos Aires,
Argentina} \\ \footnotesize{\ttfamily guerciomb@gmail.com}
\and Lisana B. Martinez \\ \footnotesize{ Instituto de Investigaciones Económicas y Sociales del Sur, Universidad Nacional del Sur, Argentina} \\
\footnotesize{Consejo Nacional de Investigaciones Cient\'ificas y T\'ecnicas (CONICET), C1033AA5, Buenos Aires,
Argentina} \\ \footnotesize{\ttfamily lisanabelen.martinez@gmail.com}}

\begin{document}
\maketitle

\begin{abstract}
Recent news cast doubts on London Interbank Offered Rate (LIBOR) integrity. Given its economic importance and the delay with which authorities realize about this situation, we aim to find an objective method in order to detect departures in the LIBOR rate that from the expected behavior. We analyze several interest rates time series and we detect an anomalous behavior in LIBOR, specially during the period of the financial crisis of 2008. Our findings could be consistent with data manipulation. 
\end{abstract}

\section{Introduction \label{sec:intro}}

London Interbank Offered Rate (LIBOR) was established in 1986 by the British Banking Association (BBA). Since its inception it has been a benchmark interest rate for pricing several derivative financial instruments traded on exchanges worldwide. Many financial contracts link parties obligations to LIBOR evolution.

The BBA defines LIBOR as ``The rate at which an individual Contributor Panel bank could borrow funds, were it to do so by asking for and then accepting inter-bank offers in reasonable market size, just prior to 11:00 London time''. Each member of the Contributor Panel
(selected banks from BBA) submits its rates every London business day through electronic means to Thomson Reuters by 11:10 a.m. London Time. When all this information is gathered, rates are ordered and the second and third quartiles are averaged. This average is published as the LIBOR rate for each maturity and currency.

On May 29, 2008 \cite{MollenkampWhitehouse} published an article on the Wall Street Journal casting doubts on the transparency of LIBOR setting, implying that published rates were lower than those implied by Credit Default Swaps (CDS). Investigations conducted by several market authorities such as US Department of Justice, and the European Commission, detected data manipulation and imposed severe fines to banks included in such illegal procedure.

The importance of a good pricing system is based on its usefulness for making decisions. As \cite{Hayek45} affirmed ``we must look at the price system as such a mechanism for communicating information if we want to understand its real function''.

In recent years, the journalist initiated investigation about LIBOR make economists to critical examine the evolution of LIBOR rates and compare it with other market rates. In this line, \cite{TaylorWilliams2009} documented the detachment of the LIBOR rate form other market rates such as the Overnight Interest Swap (OIS), Effective Federal Fund (EFF),  Certificate of Deposits (CDs), Credit Default Swaps (CDS), and Repo rates. The hypothesize that the reasons for this divergent behavior is due to expectations in future interest rates and in counterpart risk. \cite{SniderYoule} study individual quotes in the LIBOR bank panel and corroborate the claim by  \cite{MollenkampWhitehouse} that LIBOR quotes in the US are not strongly related to other bank borrowing cost proxies. In their model, the incentive for misreporting borrowing costs is profiting from a portfolio position. Consequently, the misreporting could be upward in one currency and downward in another currency, depending on the portfolio exposition. The proof of such behavior is detected with the formation of a compact cluster of the different panel bank quotes around a given point. 

\cite{AbrantesMetz2011} track the daily LIBOR rates over the period 1987 to 2008. In particular, this paper analyzes the empirical distribution of the Second Digits (SDs) of the Libor interest rate, and compares with the uniform and Benford's distributions. Taking into account the whole period the null hypothesis of that the empirical distribution follows either the uniform or Benford's distribution cannot be rejected. However, if only the period after the subprime crisis is taken into account, the null hypothesis is rejected. 

Given its extensive use, the economic consequences of a ``manipulated'' or at least a wrong LIBOR could be very important. It gives wrong signals to financial markets and general public. It could induce excessive or restricted borrowing (depending on the sense of the movement). Additionally, it gives a wrong signal to policy makers, which could not observe a credit crunch if interest rates are low

The above mentioned papers raise questions regarding the integrity of the rate quotations coming from individual banks.

The aim of this paper is to propose the use of quantifiers derived from Information Theory, in order to detect anomalies in data time series. In particular, we use permutation entropy and permutation statistical complexity, which are able to capture and discriminate different dynamics in time series. These quantifiers, when represented in a Cartesian plane conform the Complexity Entropy Causality Plane. It is as a useful visual tool to detect changes in data stochastic and chaotic dynamics and thus, possible data manipulation. Although statistical methods itself cannot establish the existence of manipulation, this tool could be useful for surveillance authorities such as central banks or exchange commissions, as an early warning device to propose in-depth investigation.

\section{Information theory quantifiers \label{sec:itq}}
We analyze time series using quantifiers based on information theory. In particular we compute two metrics: the permutation entropy and the permutation statistical complexity of each series.
Given a discrete probability distribution $P=\{p_i\}$, with $i=\{1, \dots, M\}$, Shannon entropy is defined in \cite{book:shannon1949}  as:
\begin{equation}
{\cal S}[P]= -\sum_{i=1}^M p_i \ln{p_i}
\label{eq:entropy}
\end{equation}
This quantifier estimate the average minimum number of bits needed to encode a string of symbols based on  the frequency of the symbols. It equals zero if the patterns are fully deterministic and reaches its maximum value for a uniform distribution. 
The use of informational entropy to study economic phenomena is not new. In fact, \cite{TheilLeenders65,Fama65entropy,Dryden68} could be considered as the seminal papers on this field. More recently,  \cite{Martina2011,OrtizCruz12} apply entropy and multiscale entropy analysis to assess the efficiency crude oil price. Alvarez Ramirez \textit{et al.} (2012) also use entropy methods to quantify the dynamics of informational efficiency of US stock market during the last 70 years.

However, analyzing time series only by means of Shannon entropy could be insufficient. As stressed in \cite{FeldmanCrutchfield98,FeldmanMcTague08}, an entropy measure does not quantify the degree of structure or patterns present in a process and it is necessary to introduce into the analysis a measure the statistical complexity in order to characterize the organizational properties of a system. 

A family of statistical complexity measures, based on the functional form developed by \cite{LMC95}, is defined in \cite{Martin2003,Lamberti2004119} as:
\begin{equation}
{\cal C}_{JS}= {\cal Q}_{J[P,P_e]} {\cal H} [P]
\label{eq:complexity}
\end{equation}
where ${\cal H} [P]=S[P]/S_{\max}$ is the normalized Shannon entropy, $P$ is the discrete probability distribution of the time series under analysis, $P_e$ is the uniform distribution and ${\cal Q}_{J[P,P_e]}$ is the so-called disequilibrium. This disequilibrium is defined in terms of the Jensen-Shannon divergence, which quantifies the difference between two probability distributions. \cite{paper:martin2006} demonstrates the existence of upper and lower bounds for generalized statistical complexity measures such as ${\cal C}_{JS}$ . Additionally, as highlighted in \cite{Soriano2011a}, the permutation complexity is not a trivial function of the permutation entropy because they are based on two probability distributions. A complete discussion about this measures and details about their calculation is in \cite{Zunino2010a}. 

The planar representation of these quantifiers is introduced in efficiency analysis in \cite{Zunino2010a} and was successfully used to rank efficiency in stock markets \cite{ZuninoCausality10}, commodity markets \cite{ZuninoPermutation11}, and to link informational efficiency with sovereign bond ratings \cite{Zunino2012}.

\subsection{Estimation of the probability density function \label{sec:PDF}}

In order to evaluate this quantifiers, a symbolic technique should be selected in order to obtain the appropriate probability distribution function. Following \cite{Zunino2010a,ZuninoCausality10,ZuninoPermutation11,RossoNoise07}, we use the \cite{BandtPompe02} permutation method, because it is the single symbolization technique that considers time causality. Given a time series of length $N$, this technique requires the selection of an embedding dimension (pattern length) $(D)$ and an embedding delay $\tau$. No model assumption is needed because Bandt and Pompe method makes partitions of the time series and orders values within each partition. This methodology requires only weak stationarity assumptions. The probability distribution arises naturally from pattern counting. Since we work with daily data, we select $\tau=1$. The selected pattern length should fulfill $N \gg D!$ , in order to obtain reliable quantifiers .
In order to detect changes in the generating process dynamics, we compute the permutation entropy and the permutation statistical complexity using a sliding window. We fix a window length of $N=500$ for embedding dimension $D=4$. The step between each window is $\delta=30$. We select this step because is approximately one month. We obtain a total of 118 estimation windows.
 
If a series is purely random, permutation entropy is maximized and permutation statistical complexity is minimized. Since we work with normalized quantifiers, the maximum efficiency point of the CECP is $(1,0)$. 

We argue that, without manipulation the location of each window in the CECP should be stable, or at least should not follow no predictable path.

\section{Data \label{sec:data}}

Following \cite{TaylorWilliams2009}, we select several interest rates of the UK area. In particular, we consider the LIBOR, OIS and Repo Benchmark  and Sterling OverNight Index Average (SONIA), all at 3 month maturity. The data span is from 17/05/1999 until  08/09/2014, with a total of 3996 datapoints, except for UK OIS rate that starts on 19/01/2004 (2776 datapoints). All data were retrieved from DataStream.

SONIA is  the reference rate for overnight unsecured transactions in the Sterling market, calculated as the weighted average rate of all unsecured overnight sterling transactions brokered in London by Wholesale Markets Brokers' Association (WMBA) members between 00:00 and 15.15 GMT in a minimum deal size of 25 million GBP.

Overnight Indexed Swaps are financial derivatives  used to hedge against moves in overnight interest rates 

Repurchase agreements (Repo) are collateralised lending transactions. BBA Repo Rates are calculated by a panel of 11 contributors, using an averaging method like LIBOR.

\section{Results \label{sec:results}}

Using the data described above and with the methodology described in Section \ref{sec:itq}, we compute the permutation entropy and permutation complexity for each of the series using a sliding window. The rationale for using a moving sample is to study the evolution of these quantifiers during the period under examination. Our sliding window contains $N=500$ datapoints, the frequency is daily ($\tau=1$) and the step between each window is $\delta=30$ datapoints. According to this we obtained 118 estimation windows. It is worth mentioning that we performed our analysis for embedding dimensions $D=4$. However, if we use different embedding dimensions, results are similar. We believe that, with a pattern length $D=4$ and 500 datapoints, we are able to capture the dynamics of the series under analysis, obtaining consistent estimates of both ${\cal H}_S$ and ${\cal C}_{JS}$. With the use of the sliding window, incoming perturbances are taken into account in the permutation information quantifiers.

As described in Section \ref{sec:intro}, economic theory establishes that, if prices are set in a competitive market and information is quickly embedded into prices, their time series should look as a random walk. Using the financial economics argot defined by 
\cite{Fama70}, we say that in this situation the market is informationally efficient in its weak form. This situation means that, in the context of the CECP, estimations of ${\cal H}_S$ and ${\cal C}_{JS}$ should lie on the bottom right corner. If this happens, the market is said to be informationally efficient. Otherwise, either the benchmark model is wrong or the market is informationally inefficient.

\begin{figure}[!ht]
\center
\includegraphics[scale=.99]{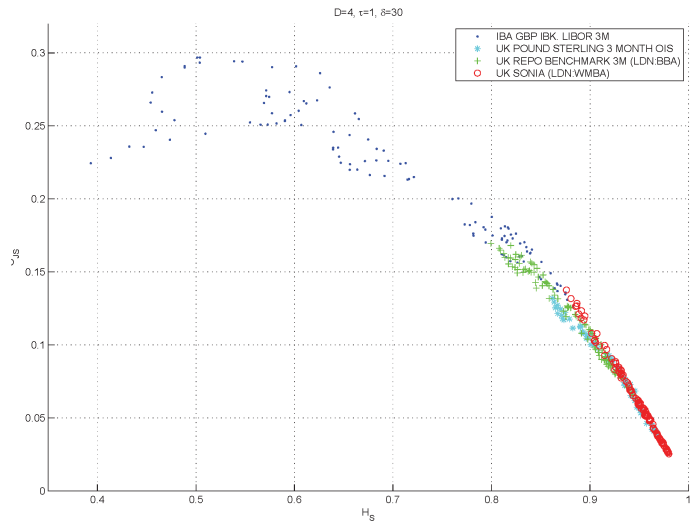}
\caption{Complexity Entropy Causality Plane for $D=4$,$\tau=1$,$\delta=30$ of UK interest rates}
\label{fig:CECP_UK}
\end{figure}

In Figure \ref{fig:CECP_UK} we display the Complexity Entropy Causality Plane of UK data. Each point reflects the calculation of permutation entropy and permutation complexity for a period of the sliding window. We observe that SONIA, Repo and OIS rates  occupy the bottom right area of the CECP. This area could be associated with random processes that exhibit no or low memory. 

\begin{figure}[!ht]
\center
\includegraphics[scale=.99]{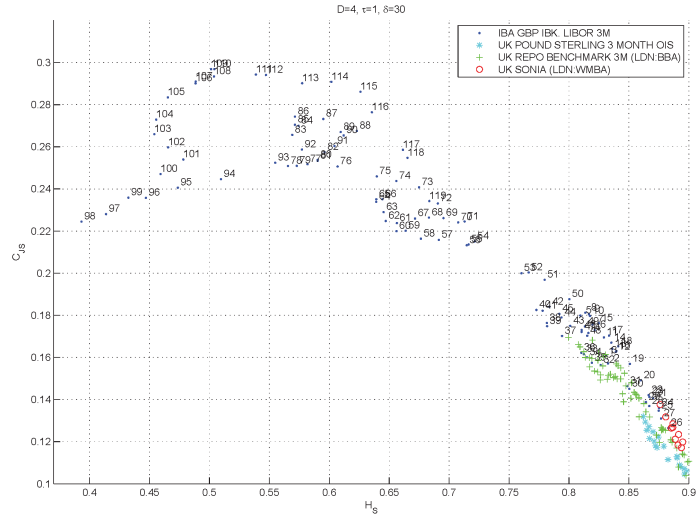}
\caption{Detail of the Complexity Entropy Causality Plane for $D=4$,$\tau=1$,$\delta=30$ of UK interest rates, focusing in the movement of LIBOR rate. Numbers refers to the estimation windows. }
\label{fig:CECP_UK_ZOOM}
\end{figure}

On contrary, LIBOR rate spreads throughout much of the CECP. In order to analyze the direction of the movement across time, we make a zoom of the CECP focusing on the LIBOR rate (see Figure \ref{fig:CECP_UK_ZOOM}). The numbers on this figure references each estimation period of this rate. We can clearly observe that LIBOR rate performs relatively good (i.e. following dynamics compatible with a random walk) until the estimation of period 53. This period spans from December 2004 until November 2006, just prior to the upheaval of the financial markets due to the subprime crisis. If we pay attention to the temporal evolution of the Information Theory quantifiers, we clearly see a path towards inefficiency areas (i.e. higher permutation complexity and lower permutation entropy). This dynamics is followed almost regularly until the estimation of period 98, which spans from February 2010 until January 2012. In May 2008 the Wall Street Journal published the seminal notice of possible LIBOR manipulation \cite{MollenkampWhitehouse}, and subsequently a series of investigations were started in order to scrutinize the behavior of individual banks in their LIBOR quotes submissions. After the estimation of period 98, the path is reverted. Our estimations reflect a slow but constant improvement in the informational efficiency of the LIBOR time series. This improvement seems to be consistent with the uncover of the LIBOR scandal and the subsequent reform in the LIBOR setting characteristics. Our methodology is not directly aimed to detect data manipulation. However, by means of it, we observe a significant change in the informational efficiency levels in the LIBOR time series, that overlaps with the landmarks of the beginning of the LIBOR manipulation and LIBOR scandal detection. In fact, several banks recognized their guilty and paid enormous fines for abusing of their market positions as published by  \cite{Liborthenandnow} and \cite{FTemail}. Putting this things together, we believe that our methodology is able to detect situations in which exogenous forces intervene in the formation of prices. 

This is specially important because the inefficiency path has not been followed by other interest rates. If LIBOR behavior is clearly different from other interest rate behavior, what is the difference between LIBOR rate and the other selected interest rates? Probably, the reason could reside in the very intrinsic characteristics of the LIBOR rate. In fact the methodology of LIBOR and the other rates differ substantially. Whereas, OIS  and SONIA reflect an average of actual transactions, LIBOR rate reflects an self-estimation of the borrowing cost of a selected group of banks and do not necessarily reflect any transaction. 

\begin{figure}[!ht]
\center
\includegraphics[scale=.99]{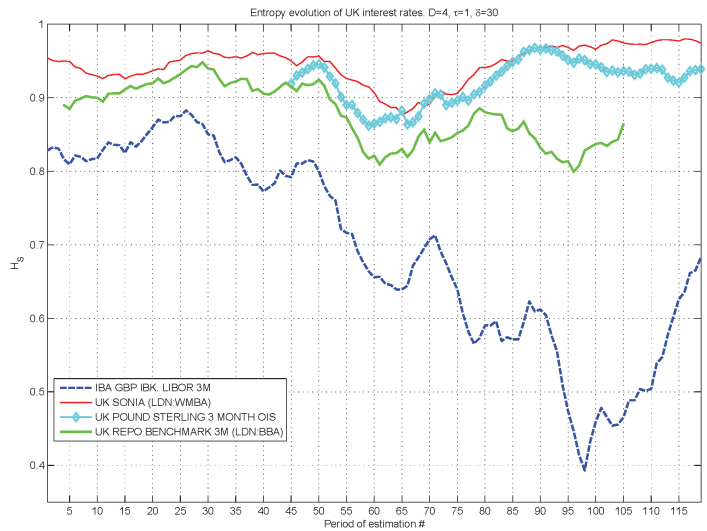}
\caption{Permutation entropy evolution of the different UK interest rates, computed for $D=4$,$\tau=1$,$\delta=30$ }
\label{fig:entro_evol_UK}
\end{figure}

In order to clarify our claim, Figure \ref{fig:entro_evol_UK} display the temporal evolution of the permutation entropy of all the interest rates. This figure clearly shows that the permutation entropy of the rates that reflect actual transactions is relatively high and their movement is bounded between 0.8 and 1. On contrary LIBOR rate exhibits a progressive deterioration in its permutation entropy, achieving a absolute minimum ($H_S\approx 0.39$) in period 98. After this point there is a steady recovery of higher levels of permutation entropy. 

\section{Conclusions \label{sec:conclusions}}
According to our results, Information Theory quantifiers are suitable instruments for studying financial market stochastic dynamics, and eventually to detect anomalies. In particular, we found that unilateral interest rates (LIBOR and Repo) exhibited a dramatic change in their underlying generating process, whereas bilateral market based rates (OIS and SONIA) were more or less constant in their stochastic characteristics. We believe that further investigation in this topic is desirable. We will continue working in the fine-tuning of our model in order to detect a possible structural break in the LIBOR generating process.

\bibliographystyle{hplain}
\bibliography{liborbib}

\begin{thebibliography}{10}

\bibitem{AbrantesMetz2011}
Rosa~M. Abrantes-Metz, Sofia~B. Villas-Boas, and George Judge.
\newblock Tracking the libor rate.
\newblock {\em Applied Economics Letters}, 18(10):893--899, 2011.

\bibitem{BandtPompe02}
Christoph Bandt and Bernd Pompe.
\newblock Permutation entropy: A natural complexity measure for time series.
\newblock {\em Phys. Rev. Lett.}, 88(17):174102, April 2002.

\bibitem{Liborthenandnow}
{Banker Middle East}.
\newblock Libor then and now...
\newblock Document BANMEA0020140519ea5j0000o, 2014.

\bibitem{Dryden68}
Myles~M. Dryden.
\newblock Short-term forecasting of share prices: an {Information Theory}
  approach.
\newblock {\em Scottish Journal of Political Economy}, 15(1):227--249, 1968.

\bibitem{Fama65entropy}
Eugene~F. Fama.
\newblock Tomorrow on the {New York} stock exchange.
\newblock {\em The Journal of Business}, 38(3):285--299, July 1965.

\bibitem{Fama70}
Eugene~F. Fama.
\newblock Efficient capital markets: A review of theory and empirical work.
\newblock {\em The Journal of Finance}, 25(2, Papers and Proceedings of the
  Twenty-Eighth Annual Meeting of the American Finance Association New York,
  N.Y. December, 28-30, 1969):383--417, May 1970.

\bibitem{FeldmanCrutchfield98}
David~P. Feldman and James~P. Crutchfield.
\newblock Measures of statistical complexity: Why?
\newblock {\em Phys. Lett. A}, 238(4–5):244--252, 1998.

\bibitem{FeldmanMcTague08}
David~P. Feldman, Carl~S. McTague, and James~P. Crutchfield.
\newblock The organization of intrinsic computation: Complexity-entropy
  diagrams and the diversity of natural information processing.
\newblock {\em Chaos}, 18(4):043106, 2008.

\bibitem{Hayek45}
F.~A. Hayek.
\newblock The use of knowledge in society.
\newblock {\em The American Economic Review}, 35(4):519--530, 1945.

\bibitem{Lamberti2004119}
P.~W. Lamberti, M.~T. Martin, A.~Plastino, and O.~A. Rosso.
\newblock Intensive entropic non-triviality measure.
\newblock {\em Physica A}, 334(1–2):119--131, 2004.

\bibitem{LMC95}
R.~L\'opez-Ruiz, H.~L. Mancini, and X.~Calbet.
\newblock A statistical measure of complexity.
\newblock {\em Phys. Lett. A}, 209(5–6):321--326, 1995.

\bibitem{paper:martin2006}
M.~T. Mart\'in, A.~Plastino, and O.~A. Rosso.
\newblock Generalized statistical complexity measures: Geometrical and
  analytical properties.
\newblock {\em Physica A}, 369:439--462, 2006.

\bibitem{Martin2003}
M.T Martin, A~Plastino, and O.A Rosso.
\newblock Statistical complexity and disequilibrium.
\newblock {\em Physics Letters A}, 311(2–3):126 -- 132, 2003.

\bibitem{Martina2011}
Esteban Martina, Eduardo Rodriguez, Rafael Escarela-Perez, and Jose
  Alvarez-Ramirez.
\newblock Multiscale entropy analysis of crude oil price dynamics.
\newblock {\em Energy Economics}, 33(5):936--947, September 2011.

\bibitem{MollenkampWhitehouse}
Carrick Mollenkamp and Mark Whitehouse.
\newblock Study casts doubt on key rate: {WSJ} analysis suggests banks may have
  reported flawed interest data for libor.
\newblock The Wall Street Journal, 2008.

\bibitem{OrtizCruz12}
Alejandro Ortiz-Cruz, Eduardo Rodriguez, Carlos Ibarra-Valdez, and Jose
  Alvarez-Ramirez.
\newblock Efficiency of crude oil markets: Evidences from informational entropy
  analysis.
\newblock {\em Energy Policy}, 41:365--373, 2012.

\bibitem{RossoNoise07}
O.~A. Rosso, H.~A. Larrondo, M.~T. Martin, A.~Plastino, and M.~A. Fuentes.
\newblock Distinguishing noise from chaos.
\newblock {\em Phys. Rev. Lett.}, 99(15):154102, Oct 2007.

\bibitem{FTemail}
Lina Saigol.
\newblock Libor: The email trail.
\newblock The Financial Times,
  http://www.ft.com/cms/s/0/cefd67a0-25df-11e3-aee8-00144feab7de.html\#axzz3FNK2XAFe,
  2013.

\bibitem{book:shannon1949}
Claude~E. Shannon and Warren Weaver.
\newblock {\em The Mathematical Theory of Communication}.
\newblock University of Illinois Press, Champaign, IL, 1949.

\bibitem{SniderYoule}
Connan~Andrew Snider and Thomas Youle.
\newblock Does the libor reflect banks' borrowing costs?
\newblock Available at SSRN: http://ssrn.com/abstract=1569603 or
  http://dx.doi.org/10.2139/ssrn.1569603, 2010.

\bibitem{Soriano2011a}
Miguel~C. Soriano, Luciano Zunino, Osvaldo~A. Rosso, Ingo Fischer, and
  Claudio~R. Mirasso.
\newblock Time scales of a chaotic semiconductor laser with optical feedback
  under the lens of a permutation information analysis.
\newblock {\em IEEE J. Quantum Electron.}, 47:252--261, 2011.

\bibitem{TaylorWilliams2009}
John~B. Taylor and John~C. Williams.
\newblock A black swan in the money market.
\newblock {\em American Economic Journal: Macroeconomics}, 1(1):58--83, 2009.

\bibitem{TheilLeenders65}
H.~Theil and C.~T. Leenders.
\newblock Tomorrow on the {Amsterdam} stock exchange.
\newblock {\em The Journal of Business}, 38(3):277--284, July 1965.

\bibitem{Zunino2012}
Luciano Zunino, Aurelio~Fern\'andez Bariviera, M.~Bel\'en Guercio, Lisana~B.
  Martinez, and Osvaldo~A. Rosso.
\newblock On the efficiency of sovereign bond markets.
\newblock {\em Physica A}, 391(18):4342--4349, 2012.

\bibitem{Zunino2010a}
Luciano Zunino, Miguel~C. Soriano, Ingo Fischer, Osvaldo~A. Rosso, and
  Claudio~R. Mirasso.
\newblock Permutation-information-theory approach to unveil delay dynamics from
  time-series analysis.
\newblock {\em Phys. Rev. E}, 82:046212, 2010.

\bibitem{ZuninoPermutation11}
Luciano Zunino, Benjamin~M. Tabak, Francesco Serinaldi, Massimiliano Zanin,
  Darío~G. P\'erez, and Osvaldo~A. Rosso.
\newblock Commodity predictability analysis with a permutation information
  theory approach.
\newblock {\em Physica A: Statistical Mechanics and its Applications},
  390(5):876--890, 3/1 2011.

\bibitem{ZuninoCausality10}
Luciano Zunino, Massimiliano Zanin, Benjamin~M. Tabak, Darío~G. P\'erez, and
  Osvaldo~A. Rosso.
\newblock Complexity-entropy causality plane: A useful approach to quantify the
  stock market inefficiency.
\newblock {\em Physica A: Statistical Mechanics and its Applications},
  389(9):1891--1901, 5/1 2010.

\end{thebibliography}

\end{document}